\documentclass[twocolumn,prl,aps,showpacs]{revtex4}
\usepackage{amssymb}
\usepackage{amsmath}
\usepackage{epsfig}

\begin{document}

\title{Application of two dimensional Frenkel-Kontorova model to nanotribology}
\author{ Wen-shan Duan$^{1,}$ }
\email[Email: ]{duanws@nwnu.edu.cn}

\author{ Cang-long Wang$^{1}$}
\author{ Xue-ren Hong$^{1}$}
\author{ Jian-min Chen$^{2}$ }

\affiliation{1.Department of Physics, Northwest Normal University, Lanzhou, 730070, P. R. China\\
2.State Key Laboratory of Solid Lubrication, Lanzhou institute of
Chemical Physics, Chinese Academy of Science, Lanzhou, 720000, P. R.
China}

\begin{abstract}
A two-dimensional Frenkel-Kontorova model is set up. Its application
to the tribology is considered. The materials and the
commensurability between two layers strongly affect the static
friction force. It is found that the static friction force is larger
between two layer of same materials than that for different
materials. For two-dimensional case the averaged static friction
force is larger for the uncommensurate case than that for the
commensurate case, which is completely different from
one-dimensional case. The directions of the propagation of the
center of mass and the external driving force are usually different
except at some special symmetric directions. The possibility to
obtain superlubricity is suggested.
\end{abstract}
 \maketitle

Driven dynamics of a system of interacting atoms is an interesting
physical problem. It has important applications in mass and charge
transport phenomena in solids and  crystal surfaces. One of
important applications has emerged in tribology studies, where a
thin atomic layer is confined between two substrates which move
respect to one another\cite{1,2}. Fundamental understanding of
friction is vitally important in many areas of science and
technology, ranging from nanotribology to crack propagation of
earthquake dynamics\cite{-1}. Progress has also been made on how to
tune the intrinsic frictional forces between a sliding subject and
the underlying substrate, through interface modification down to the
molecular or atom scale\cite{-2,-3}. Such important understanding at
the microscopic level, in turn, is expected to serve as important
guidance in the design of smart materials with desirable lubricant
properties for industrial and biomedical
applications\cite{-4,+chen}.

The application of driven Frenkel-Kontorova (FK) type model has
received an increasing interest as a possible interpretative tool to
understand the complex field of nanotribology\cite{9}. One
dimensional FK model has been extensively studied
recently\cite{13,+hu}. However, in friction systems, an importance
of higher dimensionality has been emphasized, which makes the
superlubricity (the state of vanishing friction) appear much more
easily\cite{--1,--2,+M.Hirano}. It is found in the experimental
results that by measuring the atomic-scale friction force as a
function of rotational angle between two contacting layers
superlubricity may appear in certain misfit angle between two
layers\cite{--1,--2}. Friction forces completely vanish when, for
example, incommensurately contacting surfaces sliding against each
other. In such contact, the ratio between the lattice units of the
surfaces is irrational along the sliding direction, so each
individual atom receives different amounts of force from different
directions. These forces consequently offset each other, resulting
in zero friction. This offsetting of forces is made possible by the
continuous motion of atoms, which is the basic principle behind
superlubricity. Higher dimensionality of system is, therefore,
crucial for atoms to move continuously. It is, therefore, necessary
to extend the 1D FK model to the higher dimensions to obtain a more
real model which can be realized in experiment.

For this reason, we consider an upper layer in which there are
$N\times M$ atoms and they are arranged on a 2D square lattice. We
first investigate the position and the velocity of an arbitrary
$(n,m)$th atom, where $n=1,2,\cdots, N, m=1,2,\cdots, M$. Its
position can be expressed by $\bf{r} _{n,m}=\it (x_{n,m},y_{n,m})$,
where $x_{n,m}$ and $y_{n,m}$ are the positions in  $x$ and $y$
directions respectively. For this arbitrary $(n,m)$th atom, the
interactions among the nearest and the next nearest neighbors are
considered among the upper layers. The interatomic interaction
potential is chosen to be of the simple harmonic form
%%%%%%%%%%%%%%%%%%%%%%%%%%%%%%%%%%%%%%%%%%%%%%%%%%%%%%%%%%%%%%%%%%%%%%
$V_{int}=\sum_{i,j} {K\over 2}
[(x_{i,j}-x_{n,m}-l)^2+(y_{i,j}-y_{n,m}-l)^2],$
%%%%%%%%%%%%%%%%%%%%%%%%%%%%%%%%%%%%%%%%%%%%%%%%%%%%%%%%%%%%%%%%%%%%%
with a strength $K$, a natural equilibrium spacing $l=a$ between the
nearest neighbor atoms of $(n+1,m)$th, $(n-1,m)$th, $(n,m+1)$th,
$(n,m-1)$th, and $l=\sqrt{2}a$ between the next nearest neighbor
atoms of $(n+1,m+1)$th, $(n-1,m+1)$th, $(n+1,m-1)$th, $(n-1,m-1)$th
for the upper layer.

However, this arbitrary $(n,m)$th atom not only interacts with each
other among the particles in the upper layer, but also with the
lower layer through a 2D periodic substrate potential which depends
on the lattice structure in the lower layer. For different materials
we choose different substrate potential. For generality, we choose
two kinds of substrate potentials. One is with a square lattice
symmetry,
%%%%%%%%%%%%%%%%%%%%%%%%%%%%%%%%%%%%%%%%%%%%%%%%%%%%%%%%%%%%%%%%%%%%%%%%%%%%%%%%
$V_{sqare}={f \over 2\pi}[V_0-\cos{2\pi x' \over
b}-\cos{2\pi y' \over b}]$,
%%%%%%%%%%%%%%%%%%%%%%%%%%%%%%%%%%%%%%%%%%%%%%%%%%%%%%%%%%%%%%%%%%%%%%%%%%%%%%%%
where $f$ is the magnitude of the adhesive force between the two
layers, $V_0$ is a constant and the length $b$ is the natural
equilibrium spacing of the lower layer. The other is with a
hexagonal symmetry,
%%%%%%%%%%%%%%%%%%%%%%%%%%%%%%%%%%%%%%%%%%%%%%%%%%%%%%%%%%%%%%%%%%%%%%%%%%%%%%%%
$V_{hexagonal}={f \over \pi}(V_0-\cos{2\pi x' \over b}\cos{2\pi y'
\over \sqrt{3}b})$.
%%%%%%%%%%%%%%%%%%%%%%%%%%%%%%%%%%%%%%%%%%%%%%%%%%%%%%%%%%%%%%%%%%%%%%%%%%%%%%%%

For general case of the system, in which the orientations of the two
layers do not match, we rotate the two layers with respect to each
other by an arbitrary misfit angle $\theta$. Then
%%%%%%%%%%%%%%%%%%%%%%%%%%%%%%%%%%%%%%%%%%%%%%%%%%%%%%%%%%%%%%%%%%%%%%%%%%%%%%%%%%%%%
$\begin{pmatrix}
x'\\y' \end{pmatrix}=\begin{pmatrix}  \cos{\theta} &
-sin{\theta}\\
 \sin{\theta} & cos{\theta} \end{pmatrix}
\begin{pmatrix} x\\y \end{pmatrix}
$.
%%%%%%%%%%%%%%%%%%%%%%%%%%%%%%%%%%%%%%%%%%%%%%%%%%%%%%%%%%%%%%%%%%%%%%%%%%%%%%%%%%%%%%%%%%%
The position of this arbitrary $(n,m)$th atom $\bf{r} _{n,m}$
satisfies the following equation of motion,
%%%%%%%%%%%%%%%%%%%%%%%%%%%%%%%%%%%%%%%%%%%%%%%%%%%%%%%%%%%%%%%%%%%%%%%%%%%%%%%%%%%%%%%%%%
\begin{equation}
\begin{split}
&\ddot{\bf{r}}_{n,m}+\gamma
\dot{\bf{r}}_{n,m}+{\partial(V_{int}+V_{sub}) \over
\partial\bf{r}_{n,m}} =\bf{F_{\it ext}}
\end{split}
\end{equation}
%%%%%%%%%%%%%%%%%%%%%%%%%%%%%%%%%%%%%%%%%%%%%%%%%%%%%%%%%%%%%%%%%%%%%%%%%%%%%%%%%%%%%%%%%%
where $\gamma$ is a phenomenology viscous damping coefficient. It
can be thought of representing degrees of freedom in real physical
systems which are not explicitly included in our model (e.g.,
vibrational or electronic excitations in the substrate). We use
dimensionless variables, where identical mass $m=1$ is assigned to
each atom. The frequencies of atomic vibrations are isotropic, i.e.,
$\omega_{x}=\omega_{y}=\sqrt{2\pi f}$, the characteristic time scale
is $\tau_{0}=\sqrt{2\pi /f}$. $\bf{F_{\it ext}} =\it (F_{\it
ext}\cos\alpha, F_{\it ext}\sin\alpha )$ is the external driving
force, and $\alpha$ is the angle between directions of $\bf{F_{\it
ext}}$ and the unit vector of $x$ axis.

A fourth-order Runge-Kutta algorithm was implemented to solve Eq.(1)
numerically. The averaged velocity is defined as
%%%%%%%%%%%%%%%%%%%%%%%%%%%%%%%%%%%%%%%%%%%%%%%%%%%%%%%%%%%%%%%%%%%%%%%%%%%%%%%%%%%%%%%%%%%%
${\bf {\bar{v}}}={1\over N\times
M}\sum_{i=1,j=1}^{i=N,j=M}{<\textbf{\.{x}}_{i,j}>}$,
%%%%%%%%%%%%%%%%%%%%%%%%%%%%%%%%%%%%%%%%%%%%%%%%%%%%%%%%%%%%%%%%%%%%%%%%%%%%%%%%%%%%%%%%%%%%%
where $<>$ denotes the time average. We modeled $N\times M $ atoms
of upper layer and choose the periodic boundary condition to enforce
a fixed density condition for the system. $x_{M+1}=x_{1}+Ma$,
$y_{N+1}=y_{1}+Na$. The initial condition we choose here is that the
velocity of each atom is zero and the position of each atom is at
its equilibrium position. For our system we let $N=12, M=12$. In
order to study how the static friction force varies with the
different materials of the lower layer, the numerical simulation of
the mobility as a function of the driving force and the static
friction force as a function of different system parameters of both
upper and lower layers are presented.

%%%%%%%%%%%%%%%%%%%%%%%%%%%%%%%%%%%%%%%%%%%%%%%%%%%%%%%%%%%%%%%%%%%%
\begin{figure}[h]
\begin{center}
\rotatebox{0}{\resizebox *{8.5cm}{6.5cm} {\includegraphics
{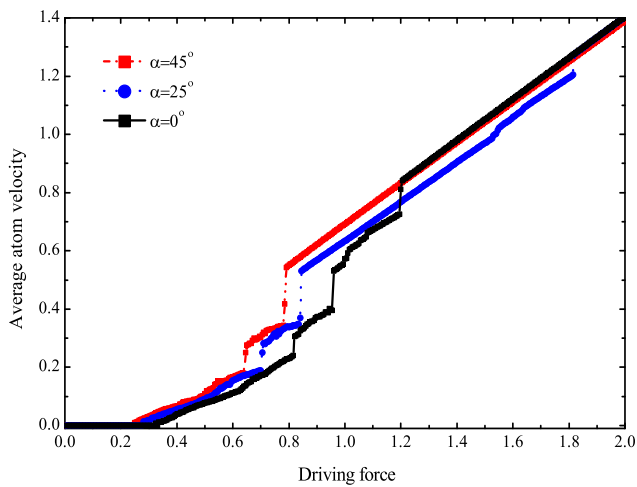}}}
\end{center}
\caption {The mobility $<{\bf v}>$ as a function of driving force
$\textbf{F}_{ext}$ for $f=1, a=b=1, m=1, K=1, \gamma=0.7$,
$\theta=40^o$ and different values of  $\alpha=0^{o}, 25^{o},
45^{o}$.}\label{trace}
\end{figure}
%%%%%%%%%%%%%%%%%%%%%%%%%%%%%%%%%%%%%%%%%%%%%%%%%%%%%%%%%%%%%%%%%%%%%%%%
%%%%%%%%%%%%%%%%%%%%%%%%%%%%%%%%%%%%%%%%%%%%%%%%%%%%%%%%%%%%%%%%%%%%%%%%%%%%%
\begin{figure}[h]
\begin{center}
\rotatebox{0}{\resizebox *{8.5cm}{7cm} {\includegraphics
{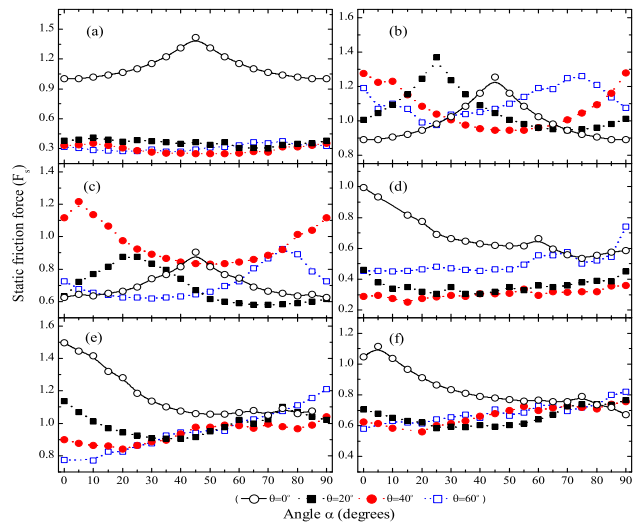}}}
\end{center}
\caption {The numerical results of $F_s$  as a function of $\alpha$
in different $\theta$. (a)The commensurate case with the square
symmetry substrate potential. (b)The golden mean case with the
square symmetry substrate potential. (c) The spiral mean case with
the square symmetry substrate potential. (d) The commensurate case
with the hexagonal symmetry substrate potential. (e) The golden mean
case with the hexagonal symmetry substrate potential. (f) The spiral
mean case with the hexagonal symmetry substrate
potential.}\label{trace}
\end{figure}
%%%%%%%%%%%%%%%%%%%%%%%%%%%%%%%%%%%%%%%%%%%%%%%%%%%%%%%%%%%%%%%%%%%%%%%%%%%%%

We define a new parameter $\beta=b/a$\cite{12,10,11}. For
simplicity, we study three typical cases of $\beta=1$ (commensurate
case), $\beta={\sqrt{5}-1\over 2}$ (golden mean case), and
$\beta=0.755 $ (spiral mean case), respectively, for both square and
hexagonal symmetric substrate potential.

Fig.1 show the numerical results  of the average  chain velocity as
a function of the driving force $\bf F_{\it ext}$  for the case of
$\theta=40^o$, $K=1$, $f=1$, and $\gamma=0.7$ with  different values
of $\alpha$. It is for the commensurate case with square symmetric
substrate potential.
%Fig.1(b) is for the case of hexagonal symmetry
%substrate potential in which $\beta=1$ as well.
%Fig.1(b) is for the golden mean case  with square
%symmetry substrate potential, and Fig.1(c) is for spiral mean case
%with square symmetry substrate potential, Fig.1(d) is for the
%commensurate case with hexagonal symmetry substrate potential,
%Fig.1(e) is for the golden mean case  with hexagonal  symmetry
%substrate potential, and Fig.1(f) is for spiral mean case with
%hexagonal  symmetry substrate potential,
We note from Fig.1 that the average velocity $\bf \bar{v}$ is zero
if the external force is less than $F_s$ (the static friction
force). As the force $\bf F_{\it ext}$ increases adiabatically, the
system undergoes a sharp transition from the pinned phase to the
running crystal phase. Our numerical results indicate that for other
values of $\theta$ the system also transfers pinning state directly
to the sliding state, although $F_s$ is different for different
$\theta$. Similar results are obtained for the case of hexagonal
symmetric substrate potential. It seems that the magnitude of $F_s$
depends on both parameters of $\alpha$ and $\theta$. Therefore,  the
static friction force depends on the external driving force
direction  and the misfit angle. We also find from our numerical
results that $F_s$ decreases as either $f$ decreases or $K$
increases.

In order to understand the anisotropic characters of the system the
numerical results of $F_s$ as a function of $\alpha$ in different
$\theta$ are given in Fig.2. It is noted that, for the case of
square symmetric substrate potential with $\beta=1$, $F_s$ strongly
depends on the parameter of $\theta$, especially at $\theta=0$ the
static friction force $F_s$ is much larger than that for other
values of $\theta$. $F_s$ also depend on the parameter of $\alpha$.
But for the golden mean case and the spiral mean case the static
friction forces $F_s$ depend on $\theta$ but not as much as that of
the commensurate case of $\beta=1$ (see Figs.2(a)-2(c)). The
variations of averaged friction forces with respect to $\theta$ are
not as much as that of commensurate case.

Fig.2 also show how the static friction force varies with respect to
different materials of the substrate. We choose two kinds of
substrate materials. One is with square symmetry and the other is
with hexagonal symmetry. We find that $F_s$ varies with different
materials of the lower layer. In order to know how the static
friction force depends on the different materials of the lower layer
we give the averaged values of $F_s$ for different substrate
potentials and different parameters of $\beta$ which is shown by
Table 1.
%%%%%%%%%%%%%%%%%%%%%%%%%%%%%%%%%%%%%%%%%%%%%%%%%%%%%%%%%%%%%%%%%%%%%%%%%%%%%%%%%%%%%%%%%%%%%%%%%%%%%%%
\begin{center}{{\textbf{Table 1 Averaged static friction forces \\
obtained from the numerical values of Fig.2.} }}
\begin{tabular}{|c|c|c|c|}\hline
  {Upper layer} & {Lower layer}  & $\beta$ & Averaged $F_s$\\
  \hline
                 &                 & 1 & 0.521\\
  \cline{3-4}
                 & Square lattice  & 0.755 & 0.766\\
  \cline{3-4}
  Square lattice &                 & 0.618 & 1.058\\
  \cline{2-4}
                 &                 & 1 & 0.46\\
  \cline{3-4}
                 & Hexagonal lattice  & 0.755 & 0.713\\
  \cline{3-4}
                 &                    & 0.618 & 1.013\\
  \hline
\end{tabular}
\end{center}
%%%%%%%%%%%%%%%%%%%%%%%%%%%%%%%%%%%%%%%%%%%%%%%%%%%%%%%%%%%%%%%%%%%%%%%%%%%%%%%%%%%%%%%%%%%%%%%%%%%%%%%

It seems from Table 1 that the static friction force $F_s$ is larger
between two layers of same materials than that for different
materials which is in agreements with experimental
results\cite{chen2}.

Meanwhile, we note from Fig.2 that the averaged friction forces for
two typical uncommensurate cases are larger than that of the
commensurate case of $\beta=1$ for both square  and hexagonal
symmetric substrate potentials. It means that for 2D case the
averaged friction force is larger for the uncommensurate case than
that for the commensurate case. This result is completely different
from 1D case, in which case the friction force is larger for the
commensurate case than that for the uncommensurate
case\cite{+M.Hirano,12}.
%The numerical results show that the system parameters, such as the
%external driving force,  plays a crucial role in the scenario of the
%lock-to-sliding transition and determination of the structure of the
%sliding state.

Now we analyze the atom trajectories for the case of $f=1,
\gamma=0.7, a=1, b=1, K=1, \theta=40^o, $ and $\alpha=25^o$ with the
square symmetric substrate potential shown in Fig.1. As the external
driving force increases, each atom of the system may move from its
equilibrium position. Fig.3(a) present the trajectories of the
$(5,4)$th, $(5,5)$th and $(5,6)$th atoms at $F_{\it ext}=0.1$. It
seems that there are small displacements around their equilibrium
positions for these atoms. However, the displacements are much less
than lattice spacing $a$. Each atom move near its equilibrium
position. However, the averaged atom velocity is still approximately
zero. This state corresponds to a pinned state. When the driving
force increases, the state will be transferred into a sliding state
of moving crystal, see Fig.3(b) for $F_{\it ext}=1.82$, and Fig.3(c)
for $F_{\it ext}=1.84$, respectively. Fig.3(b) show that the
direction of average atom velocity is not same as that of the
external driving force. The propagation direction of each atom is
not a constant, but there are vibrations around its average
direction. It actually presents a atom motion in a solitonlike
fashion. We observe a formation of moving kinks for each atom. It
suggest that for different driving forces the directions of average
atom velocity are different. Fig.3(c) show a complete crystalline
state in which case the directions between driving force and the
average atom velocity are same. Although there are small vibrations
between atom trajectory and the direction of the driving force, it
becomes smaller and smaller as the driving force increases.
%%%%%%%%%%%%%%%%%%%%%%%%%%%%%%%%%%%%%%%%%%%%%%%%%%%%%%%%%%%%%%%%%%%%%%%
\begin{figure}[h]
\begin{center}
\rotatebox{0}{\resizebox *{8.5cm}{6.5cm} {\includegraphics
{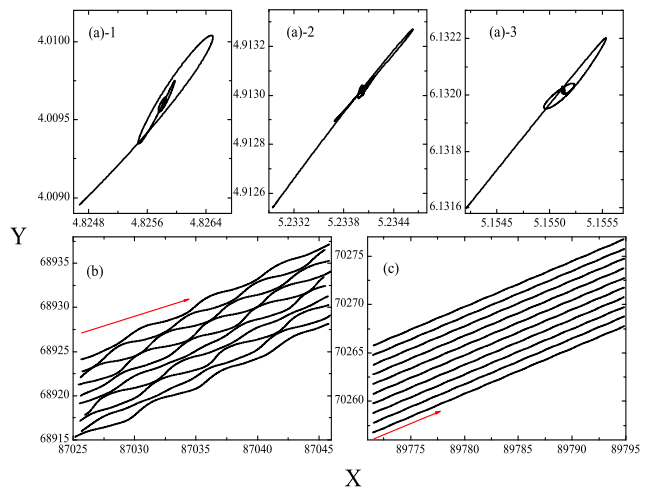}}}
\end{center}
\caption {The trajectories of selected atoms for different external
driving forces. (a) The trajectories of the atoms of $(5,4)$th,
$(5,5)$th and $(5,6)$th  at $F_{\it ext}=0.1$. (b) The trajectories
of the atoms of $(5,1)$th, $(5,2)$th, $(5,3)$th,  $(5,4)$th,
$(5,5)$th,   $(5,6)$th,   $(5,7)$th,   $(5,8)$th,   $(5,9)$th,
 and $(5,10)$th at $F_{\it ext}=1.82$, (c)  The trajectories
of the atoms of $(5,1)$th, $(5,2)$th, $(5,3)$th,  $(5,4)$th,
$(5,5)$th,   $(5,6)$th,   $(5,7)$th,   $(5,8)$th,   $(5,9)$th,
 and $(5,10)$th at
$F_{\it ext}=1.84$. }\label{trace}
\end{figure}
%%%%%%%%%%%%%%%%%%%%%%%%%%%%%%%%%%%%%%%%%%%%%%%%%%%%%%%%%%%%%%%

Fig.3 only show the trajectories of particular atoms of the system.
In order to know the trajectory of  mass center (c.m.) of the whole
system, the differences  between the direction of external driving
force and the direction of trajectory of the center of mass are
given in Fig.4. We define a parameter of $\alpha'$ to represent the
angle between directions of trajectory of c.m. and the unit vector
of $x$ axis. The dependence of $\alpha'$ on the external driving
force are shown for $\theta=0^o, 45^o, 40^o$ in Fig.4(a), 4(b),
4(c), respectively. It is found that the propagation direction of
the center of mass is same as that of the external driving force
when $\alpha=45^o$ for both $\theta=0^o$ and $\theta=45^o$ due to
the symmetry of the system. However, two directions are different
for other cases, even for the cases of $\theta=0^o$ and
$\theta=45^o$. For other values of $\theta$ two directions are
different for any values of $\alpha$. We also note that if the
external driving force is below the static friction force, although
it is not in sliding case, the propagation direction of the center
of mass is always in the direction of $\alpha'=45^o$.
%%%%%%%%%%%%%%%%%%%%%%%%%%%%%%%%%%%%%%%%%%%%%%%%%%%%%%%%%%%
\begin{figure}[h]
\begin{center}
\rotatebox{0}{\resizebox *{7.5cm}{10cm} {\includegraphics
{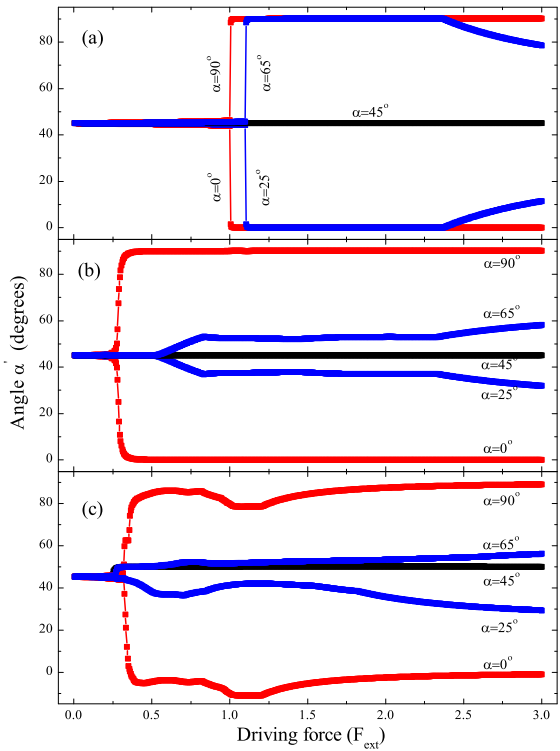}}}
\end{center}
\caption {The angle $\alpha'$ between the direction of trajectory of
the center of  mass and unit vector of $x$ axis as a function of the
magnitude of the external driving force $F_{ext}$. (a)$\theta=0^o$.
(b)$\theta=45^o$. (c)$\theta=40^o$.}\label{trace}
\end{figure}
%%%%%%%%%%%%%%%%%%%%%%%%%%%%%%%%%%%%%%%%%%%%%%%%%%%%%%%%%%%%%

In a conclusion, we find that the averaged static friction force is
larger between two layers of same materials than that for different
materials. For 2D case the averaged friction force is larger for
uncommensurate case than that for commensurate case.  To obtain
superlubricity, we may choose  different materials, but with
commensurate ratio between two layers with larger stiffness strength
$K$ of upper layer  and smaller magnitude of adhesive force between
two layers $f$, which may be a case of real system such as
diamond-like carbon film\cite{+chen}, about which superlubricity may
be realized. The directions of the propagation of the center of mass
and the external driving force are usually different. We may devise
a experiment to verify this result in the future and it may be
useful to the many fields of condensed physics, such as vortex
lattices in superconductors\cite{1,2}, Josephson junction, charge
density waves(CDW)\cite{3}, colloids\cite{4}, Wigner
crystal\cite{5}, metallic dots\cite{6,7}, magnetic bubble
arrays\cite{8}, etc.

The authors are grateful to the National Natural Science Foundation
of P. R. China (Grant No. 50575217, 10875098 and 50421502), the
Natural Science Foundation of Northwest Normal University (Grant No.
NWNU-KJCXGC-03-17).

\end{document}